\documentclass{llncs}

\usepackage{graphicx} 
\graphicspath{{./figures/}}

\usepackage[export]{adjustbox}
\usepackage{amsmath,amssymb,amsfonts,color} 
\usepackage{hyperref}
\usepackage{booktabs}

\usepackage{tabularx,ragged2e} 
\newcolumntype{L}{>{\RaggedRight\arraybackslash}X}
\definecolor{orange}{rgb}{1,0.5,0}

\usepackage{numprint}

\begin{document}

	\title{
		Towards a Semantic Search Engine for \\Scientific Articles
	}
	
	\titlerunning{Towards a Semantic Search Engine for Scientific Articles}

	\author{Bastien Latard\inst{1,2} \and Jonathan Weber\inst{1} \and \\ Germain Forestier\inst{1} 
		\and Michel Hassenforder\inst{1}} 
	\institute{MIPS, University of Haute-Alsace, Mulhouse, France 
		\and MDPI AG, Basel, Switzerland 
	}
	\authorrunning{Bastien Latard et al.}
	
	\maketitle

\begin{abstract}
Because of the data deluge in scientific publication, finding relevant information is getting harder and harder for researchers and readers.
Building an enhanced scientific search engine by taking semantic relations into account poses a great challenge. As a starting point, semantic relations between keywords from scientific articles could be extracted in order to classify articles. This might help later in the process of browsing and searching for content in a meaningful scientific way. 
Indeed, by connecting keywords, the context of the article can be extracted.
This paper aims to provide ideas to build such a smart search engine and describes the initial contributions towards achieving such an ambitious goal. 
\end{abstract}

\section{Introduction}
\label{intro}
Keeping up-to-date in a specific research field is a tedious and complex task. 
This is mandatory as it allows researchers to increase their knowledge on a domain and acquire latest ideas.
%
%
Hence, choosing the correct approach is the first step of any research work.
Despite---\textit{or because of}---the data deluge in scientific publication, 
%
%
researchers spend a significant amount of time searching for articles related to their scientific interests.

An editorial from \textit{Nature}~\cite{nature_editorial_gold_2012} clearly expressed the continued frustration of the scientific community concerning the incredible potential that text mining of scientific literature represents.
However, text miners often face the barrier of publishers' legal restrictions (i.e., closed access). 
The average growth of scientific literature is estimated to be 3 million new articles per year from journals and conferences over the last 4 years, with 3.3 million articles produced in 2016 (\url{http://www.scilit.net}).
This massive amount of data is published by more than 6000 publishers in around 47,000 scientific journals. These de-centralised and separated platforms further complicate the research process because scientists are unable to go through them all in order to search for relevant articles. Thus, they have to rely on big databases or indexing companies 
which provide either an incomplete corpus due to selection criteria or 
only display articles from their own platforms. Moreover, their search engines often offer very limited search functionalities, and this is the problem we want to tackle. 

To tackle this problem, our approach consists in using semantic relations between keywords to extract the main categories of the articles. This approach simultaneously validates both the context of the article and the context of the word, thus providing the correct category. 
Effendy and Yap~\cite{effendy_problem_2016} discussed the potential of using semantic mining tools to extract the best category of a conference. This is exactly what our framework aims to do.

\section{Method}

\label{approach}

Our approach uses BabelNet~\cite{navigli_babelnet:_2012} which is a multilingual lexicographic and encyclopaedic database based on the smart superposition of semantic lexicons (WordNet
, VerbNet) together with other collaborative databases (Wikipedia and other Wiki data).
A query for a term through BabelNet returns "dictionary entries", synonyms, categories or domains. Each synset $S$ contains the relative categories $C$, domains $D$ and synonyms $syn$ within the specific concept: 

\begin{equation}
\label{eq:synset}
S = \{C, D, syn\}
\end{equation}

\vspace{0.5em}
Assuming that synonyms of keywords might be an interesting way to connect several articles, BabelNet is the knowledge database on which our framework will rely. However, BabelNet lacks specificity and searching for one word can return synsets from various different contexts. For example, "flight" returns 36 synsets, from a South Korean movie to the verb 'to fly'. Consequently, a method to filter out unrelated synsets is mandatory.

Because synonyms are too specific, and domains are too general, categories have been naturally chosen in order to identify overlapping between synsets from different keywords. 
Indeed, if several keywords share the same category, then this is potentially the correct category in regards to the article context. In addition, the greater the number of keywords sharing the same category, the higher the confidence.
Thus, connecting the returned synsets based on their categories is an interesting way to naturally filter out all of the unrelated synsets.

This approach does filter some content, but still returns "living people; English-language films; celestial mechanics; American films" as the main categories for keywords "nonlocal gravity; celestial mechanics; dark matter". Constant noise (\textit{*\_singer}, \textit{*\_album}, etc.), meaningless in our scientific context, has been identified. A parameter can now be set in order to force the automatic filtering of identified noise.
Most of the remaining noise is finally naturally filtered out, and "celestial mechanics" is finally returned as the main category. 

Our final goal is to apply this valuable added knowledge to all articles from the scientific literature database, Scilit (\url{http://www.scilit.net}), developed by MDPI (\url{http://www.mdpi.com}).
To validate our approach, a manual analysis on a subset of 595 articles from seven journals (six about Physical Science and one about Pediatrics) has been conducted. 
We evaluated the correctness of the categories based on the connection of keywords by their synsets. 
This approach provides good precision---from 96\% to 100\%---depending on the threshold which identify the data as correct not. Indeed, strictly selecting only categories 
shared by three different keywords or more leads to a high degree of confidence (100\% precision), but a recall of 9\%. By being more tolerant and considering all categories shared by at least two keywords, precision slightly decreased (96\%) but we significantly gain in recall (47\%). Moreover, similar proportions are observed for \textit{Children}, the journal about Pediatrics (from 100\% to 92\%). This validates that our approach may be used in several domains.

\begin{figure}[]
	\includegraphics[width=0.85\textwidth,center]{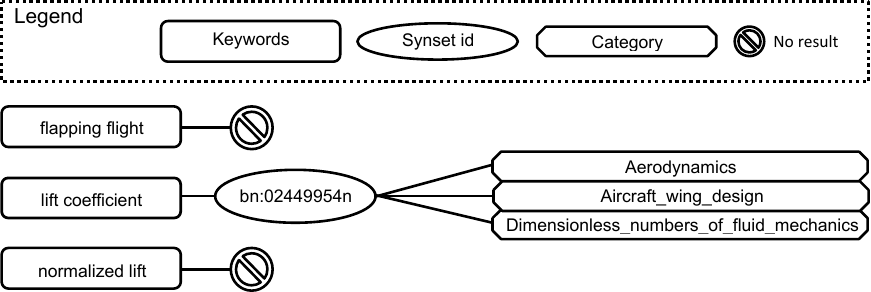}
	\caption{Limits of the exact search: only one keyword from three return data.}
	\label{fig:step1}
\end{figure}

The main drawback of our approach is that correct categories have been identified for only 22\% of the articles within the subset. 
Figure~\ref{fig:step1} illustrates the reason for the law recall and coverage of our approach.

One of the reasons for this law coverage is that BabelNet often returns no result for composed keywords (multi-word keywords), as shown in Figure~\ref{fig:step1}, where no data is returned for two of three keywords. 
In our approach---proposing only categories shared by at least two keywords---the degree of confidence is not high enough to return the categories. 
We will investigate further a way to propose some categories from these composed keywords in our future work. In doing so, we aim to significantly gain in recall, and cover many more articles.

\section{Perspectives}
Making scientific recommender systems smarter is crucial in order to help scientists in their mandatory and tedious bibliographical research phase. 
The approach proposed could be the first step in building such a smart system.
Indeed, analysing the correctness of the main category based on the overlapping of the keywords category confirms the logic of our approach. In the future, we plan to extend the search in order to extract categories from composed keywords. Splitting on spaces 
would provide some data for sub-keywords. Then, applying the same logic as described in our approach (i.e., connect by common category) will filter out unrelated items, and categories from connected items might be used for the global category connection.
By taking the example from Figure~\ref{fig:step1}, splitting "flapping flight" on spaces will return 3 and 25 synsets, respectively for "flapping" and "flight":
\begin{figure}[h]
	\centerline{\includegraphics[width=0.6\textwidth]{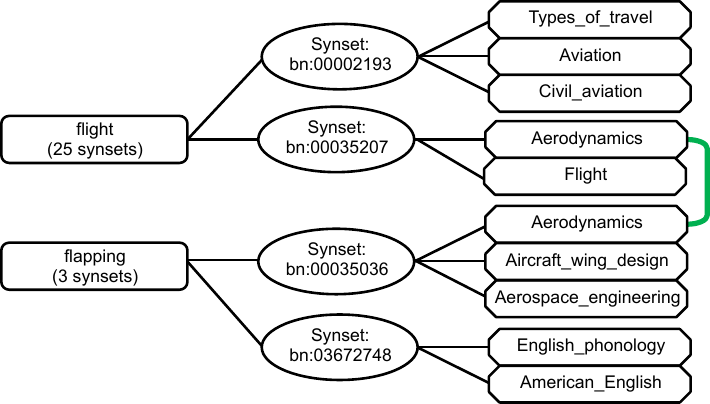}} 
	\caption{The category "Aerodynamics" is returned as the main category of "flapping flight". Other categories are filtered out by this connection.}
	\label{fig:step2}
\end{figure}

\begin{figure}[h]
	\includegraphics[width=\textwidth]{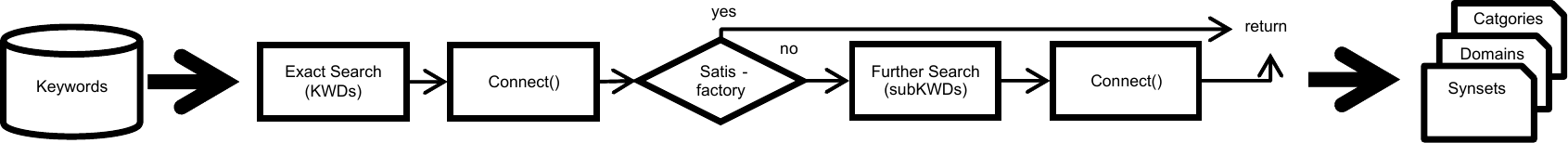}
	\caption{Illustration of the general logic of our approach in a future work}
	\label{fig:general_diagram}
\end{figure}

This further search will successfully identify "Aerodynamics" as the main category of "flapping flight". Thus, our approach would connect "Aerodynamics" based on both keywords.
Extracting the part-of-speech (with a syntactical analyser like SyntaxNet \cite{andor_globally_2016} or CoreNLP \cite{manning_stanford_2014}) from long keywords could be an interesting extra source of information for refining requests on BabelNet. 
%
Finally, Figure~\ref{fig:general_diagram} shows the main logic of our next contribution: to process in a smarter way the keywords that do not return any satisfactory results.
Later, we might also generate a graph inherited from the BabelNet's synsets as in~\cite{franco-salvador_cross-domain_2015}

\bibliographystyle{splncs}
\bibliography{zotero_23_June_2017}

\end{document}